\documentstyle[aps,tighten,12pt,epsfig,prb]{revtex}

\newcommand {\bi} {\bibitem}
\newcommand {\be} {\begin{equation}}
\newcommand {\beq} {\begin{eqnarray} \nonumber }
\newcommand {\ee} {\end{equation}}
\newcommand {\eps} {\epsilon}
\newcommand {\ei} {\epsilon_i}
\newcommand {\ej} {\epsilon_j}
\newcommand {\ek} {\epsilon_k}
\newcommand {\epp} {\epsilon^{\prime}}
\newcommand {\epd} {\epsilon^{\prime \prime}}
\newcommand {\si} {\sigma}

\newcommand {\al} {\alpha}
\newcommand {\aeps} {\alpha_{\epsilon}}

\newcommand {\app} {\alpha_+}
\newcommand {\amm} {\alpha_-}

\newcommand {\Tr} {\mbox{Tr}}

\def \form#1 {eq. (\ref{#1}) }
\def \parziale#1#2  {{\partial {#1} \over \partial {#2}}}
\newcommand{\bq}{\begin{eqnarray}}
\newcommand{\eq}{\end{eqnarray}}
\newcommand{\n}{\noindent}
\newcommand{\nn}{\nonumber\\}

\newcommand{\simg}{\stackrel{>}{\sim}}
\newcommand{\bc}{\begin{center}}
\newcommand{\ec}{\end{center}}

\newcommand {\bb}{\beta}

\newcommand {\s} {\sigma}

\newcommand {\w} {\omega}

\def\(({\left(}
\def\)){\right)}
\def\[[{\left[}
\def\]]{\right]}
\def\bi{\bibitem}

\def \(({\left(}
\def \)){\right)}
\def \[[{\left[}
\def \]]{\right]}

\def \nn{\nonumber}

\def \tk{T_K}
\def \teff{T_{eff}}
\newcommand {\la} {\langle}
\newcommand {\ra} {\rangle}

\begin{document}


\title{Lennard-Jones binary mixture: a thermodynamical
approach to glass transition}
\author{Barbara Coluzzi$^{a}$,
Giorgio Parisi$^{b}$ and Paolo Verrocchio$^{b}$}

\maketitle

\begin{center}

a) John-von-Neumann-Institut f\"ur Computing (NIC) \\
c/o Forschungszentrum J\"ulich, D-52425 J\"ulich (Germany) \\

b) Dipartimento di Fisica and Sezione INFN,\\
Universit\`a di Roma ``La Sapienza'',
Piazzale Aldo Moro 2,
I-00185 Rome (Italy)


\end{center}

\vspace{2cm}

\begin{abstract}
\noindent We study the liquid-glass transition of the Lennard-Jones binary
mixture introduced by Kob
and Andersen from a thermodynamic point of view.  By means of the replica
approach, translating the
problem in the study of a molecular liquid, we study the phase transition due
to the entropy crisis
and we find that the Kauzmann's temperature $\tk$ is $\sim 0.32$.  At the end
we compare analytical
predictions with numerical results.
\end{abstract}



\newpage

\begin{section}{Introduction}
\noindent The aim of this paper is to study the liquid-glass transition and
investigate the
glass phase of a binary mixture of particles which describes a fragile liquid.
 We consider a binary mixture interacting via a Lennard-Jones
potential, introduced by
Kob and Andersen \cite{KoAn}.  This model is suitable for such
problems because of
the lack of a crystalline state.  This allows a detailed comparison between
analytical results and
numerical simulations.

The proposed theory relies on the following assumptions,
discussed in detail in previous papers \cite{MePa1,Me,MePa2,sferesoft}:

\begin{itemize}

\item
In the supercooled liquid phase, the phase space can be partitioned in an
exponentially high number of potential energy minima separated by
energy barriers of order $O(1)$ (the barriers will diverge at a
low temperature, at the Kauzmann temperature where the viscosity diverges).
The basins of attractions of these minima are the so-called
'inherent structures' \cite{St}. A growing number of
numerical studies are supporting this picture \cite{SaDeSt,BhBrKrZi,AnPaRuVi,ScTa}.
Choosing the thermodynamic point of view, we focus mainly on
free-energy landscape. We assume that below the dynamical (Mode
Coupling \cite{GoSj}) critical temperature, $T_D$ \cite{FrPa1,CaFrPa}, the
supercooled liquid at equilibrium is almost always trapped in one of the
exponentially large number of local free energy minima, which we will call
'valleys'.  This picture is very similar to the one which
describes generalized mean field models of spin-glasses displaying a one step
replica symmetry breaking transition
\cite{KiThWo,CuKu1,crisomtap,ACP}. The free-energy minima in short
range systems are not separated by infinite barriers, so different
valleys with the same free energy can be explored by the system during its
evolution. This is what is supposed to happen in the liquid phase.

We write the partition function as a sum of contributions from different
minima:

\begin{equation}
\label{parti}
Z = e^{-N \beta \phi(T)}= \int \: df \: e^{-\beta N \: f}{\cal N}(f),
\end{equation}
where ${\cal N}(f,T)$ is the
number of minima of the free energy at temperature $T$ as function of the free
energy density $f$.
We can show that the system does not choose the configuration with minimal
free-energy, but instead minimizes the generalized free energy
density, namely:

\begin{equation}
\label{entroconfigu}
\phi(T) \sim \min_{f} [ f - T \Sigma(f,T)].
\end{equation}
The total entropy density is given by the sum of the entropy density
of a typical minimum and of the complexity, or configurational entropy
$\Sigma \sim \log {\cal N}/N$, taking into account
the number ${\cal N}$ of these minima.

\item
In the glass phase, the system is described, to a first approximation, as an
amorphous solid, where the only degrees of freedom are small displacements
around disordered centers of oscillation. We assume further that
thermodynamic quantities are self-averaging with respect to the disorder.

\end{itemize}

The main hypothesis relies on old Adam-Gibbs-Di Marzio scenario
\cite{Ka,GiDi,AdGi,Pa1a} where
the liquid-glass thermodynamic transition is related to an 'entropy crisis',
that is the existence
of a finite temperature where ergodicity breaks and the configurational
entropy vanishes.  Within
this picture, the transition temperature, $T_K$, corresponds to the temperature
where the viscosity
diverges with a generalized Vogel-Fulcher law \cite{An}, $\eta \propto \exp
(T-\tk)^{-\nu}$.  In
other words, the typical singular behaviour of experimental measures of
thermodynamic quantities
(e.g. specific heat) at a somewhat arbitrary $T_g$ is related at a 'true'
second order transition
which occurs at a lower temperature. The internal energy is continuous at
$T_K$, there is no latent
heat and the specific heat suddenly falls from the liquid value to a
definitely smaller one, in
agreement with the Dulong and Petit law.  Nevertheless, from the point of view
of an appropriate order
parameter, we find that the transition is discontinuous.  Indeed, while in the
glass phase each particle
is confined in a 'cage' of finite size due to interactions with its neighbors,
in the liquid,
because of diffusion, there is no cage, i.e. there is no thermodynamically
stable cage, but only
some metastable 'caging' effect.

The glass phase is studied by generalizing a recently proposed theory
\cite{MePa1,MePa2} to binary mixture, allowing one to obtain
equilibrium properties of fragile glasses \cite{An} from
those of liquid phase, computed for a molecular liquid consisting of $m$
'clones' of the system with \cite{Me}  $m < 1$.  This extension has
been accomplished so far only for a  soft sphere binary
mixture model \cite{sferesoft}, while in the present paper we address the
more realistic model
of a Lennard-Jones binary mixture. Due to the different nature of the
system and to the  presence of a liquid-gas transition, the simple
Hyper-Netted-Chain \cite{Han1} approximation previously used 
\cite{sferesoft} is
not enough precise. We must introduce some slightly different
approximation in investigating liquid and glass phases which will be discussed
in detail later.

After first introducing the model (section II), we present and discuss
approximations needed to compute replicated free-energy
(sections III-IV), we then describe a way to study supercooled liquid
phase (section V) and finally we present analytical results and
comparisons with numerical simulations (VI).
\end{section}

\begin{section}{Lennard-Jones binary mixtures}

We study mixtures of two types of particles with different radii, here called
$+$ and $-$, with pairwise interactions.  The Hamiltonian of our problem is:

\begin{equation}
H=\sum_{1 \le i < j \le N} V^{\epsilon_i \epsilon_j}(x_i-x_j) \ \ \ \
\epsilon_i \in \{-,+\},
\label{zn1}
\end{equation}

\noindent
where the particles move in a volume $V$ of a $d$-dimensional space,
and $V^{++},V^{+-}$, and $V^{--}$ are arbitrary short range potentials.

The so-called soft sphere (SS) model, where the potentials $V(r)$ are
short ranged and purely repulsive (i.e. proportional to $R^{-12}$),
was studied in a previous work \cite{sferesoft}. In that paper one:

\begin{itemize}
\item
demonstrated the feasibility of extending the thermodynamic
theory of liquid-glass transition recently presented \cite{MePa1,MePa2}
to binary mixtures;
\item
compared analytical results to numerical simulations;
\item
compared different methods of approximation proposed
\cite{MePa1,MePa2} to compute 'molecular' or replicated free-energy.
\end{itemize}

The free energy in the uncorrelated liquid phase and in the
correlated glass phase were there computed \cite{sferesoft} using
the framework of HNC closure \cite{Han1}. By the HNC
approximation the computed entropy in liquid phase was found correct
to within 10 \%. Other approximations were introduced to compute the
entropy in glass phase, namely the 'quenched' approximation, which
neglects the feedback of vibrational modes on center of masses,
and the 'superposition' approximation, where $p$-point correlation
functions are expressed as 'chain' products of two-point
correlation functions. When more realistic potentials are introduced,
it is sometimes necessary to use more sophisticated integral equations
than those based on the HNC.

In this paper we study a model of binary mixture introduced by Kob and
Andersen \cite{KoAn}, where potentials are of the Lennard-Jones (LJ) type:

\begin{equation}
V^{\eps\epp}(r)= 4 \eta^{\eps \epp}
\left[ \left ( \frac{\sigma_{\eps \epp}}{r} \right )^{12}
-\left ( \frac{\sigma_{\eps \epp}}{r} \right )^{6} \right].
\label{model}
\end{equation}
The concentrations used are $c_+ = {4 \over 5}$ and $c_- = {1 \over 5}$.
With the choice of parameter values $\s_{++} = 1.0$,
$\s_{+-} = 0.8$, $\s_{--} = 0.88$,
$\eta_{++}=1.0$, $\eta_{+-}=1.5$ and $\eta_{--}=0.5$
it is possible to prevent both
crystallization and separation in two phases with particles of different type.
For the sake of clarity we underline that the symbols $+$ and $-$ correspond
to $A$ and $B$ of the Kob and Andersen paper and, more
importantly, here we do not use any cut-off on the potentials.

The main effect of the introduction of an attractive term in potential
$V(r)$ is the appearance of a density-dependent temperature,
$T_{GL}(\rho)$, where the system undergoes a separation into
a liquid and gaseous phases with different densities.

This model has been extensively studied numerically in the past
\cite{KoAn,KoDoPlPoGl,DoDouKoPlPoGl,DoGlPoKoPl,SaDeSt,BaKo}. In
particular, at density $\rho = 1.2$, it has been shown
that a 'dynamical' transition occurs at temperature $T_D = 0.435$.
Below this temperature aging phenomena \cite{CuKu1,CuKu2,CuKu3}
and violation of the fluctuation-dissipation theorem \cite{BaKo,MuRi} are
observed.

In this paper, the same model is studied from a 'thermodynamic'
point of view, showing the existence of a temperature,
$\tk = 0.32 \: (< T_D)$, where the system undergoes a real
liquid-glass transition.

As a preliminary step, let us note that we have to go beyond HNC
approximation to study the model in the fluid phase.
Indeed, within this closure, integral equations at density $\rho =1.2$
show a singularity at $T_{GL} \sim 0.5$, overestimating the true $T_{GL}$.
We are forced, then, to introduce some more reliable approximation in
order to study this model.
The correlation functions used in this paper are computed solving
an integral equation which continuously interpolates between the
HNC approximation at long distances and the Mean Spherical Approximation
(MSA) at short distances, as proposed by Zerah and Hansen \cite{ZeHa}.
The details are described below.

\end{section}

\begin{section}{General framework}

The basic ideas used to study the glassy phase are the ones discussed
in the introduction and in previous papers \cite{MePa1,Me,MePa2,sferesoft}.
We briefly review here how to use them for explicit computations.

Letting $f_{eq}$ be the equilibrium free energy, which minimizes
$\phi$, the system enters the glass phase when the complexity
$\Sigma(f_{eq},T)$ vanishes. Therefore the computation of $\Sigma(f,T)$
it is the first fundamental topic of our investigation.

This issue can be settled by using the method proposed by Monasson \cite{Mo}, where $m$
identical replicas of the original system are considered.  By introducing a small but extensive and
attractive coupling term among them, these replicas are forced to be in the
same valley, so that they become strongly correlated.
In the limiting case of no coupling term, two arbitrary replicas in
the liquid phase are always found in two different valleys, because of
the exponentially large number of
them, which makes the probability of finding any two systems in the
same valley negligible.  The
transition to glass phase is supposed to be related to an ergodicity breaking,
where the
free-energy barriers become infinitely large.  In the glass phase replicas
remain correlated
even after sending the coupling term to zero.  In a sense, the coupling term
acts like, for example,
the external magnetic field in the Ising model, exposing the transition.
Moreover, the crucial point
is that using the same argument as for (\ref{entroconfigu}) the free energy of the
replicated system $\Phi(m,T)$ is related to the complexity $\Sigma(f,T)$ by
the simple formula:

\begin{equation}
\Phi(m,T) \sim \min_f [m f(T) - T \Sigma(f,T)],
\end{equation}

It is clear from this formula that if we are able to compute this
thermodynamic
potential for an arbitrary, non-integer value of the parameter $m$, we can
obtain all the values
available for free energy, $f$, and complexity, $\Sigma$, simply by varying this
parameter $m$ in the
following formulas:

\begin{equation}
\label{freecomp}
{\partial \Phi(m,T) \over \partial m} =f \hspace{.5cm}
{m^2 \over T} {\partial (\Phi(m,T)/m) \over \partial m}=\Sigma.
\end{equation}

By eliminating $m$ from these formulas one obtains the complexity
as a function of the
free energy $\Sigma(f)$.  Obviously, the physical free-energy and
configurational entropy are
obtained in the limit $m \to 1$.

Because the function $\Phi(m,T)/m$ should be a convex function of parameter $m$, a
physically possible situation is
described only in the interval $m \in [0,m^*]$, where $m^*$ is the point of
maximum.  Indeed,
$m^*$ corresponds to the minimum value $f_{min}$ available for free-energy.
For greater values of
$m$, the second equation in (\ref{freecomp}) implies that complexity is
negative, which cannot be accepted on physical grounds.

The region of temperatures where $m^* >1$ is the liquid phase, where equilibrium
complexity is positive
and the free energy is that which minimizes the balance between $\beta f$
and $\Sigma(f,T)$.  In
the low temperature glass phase, $m^*<1$ instead, and in the whole region $m
\in [m^*,1]$ the
replicated free energy (per replica) has the value $\Phi(m,T)/m = f_{min}(T)$.
 In particular we
have that the equilibrium free energy $\Phi(1,T) = f_{min}(T)$.
From this free-energy it
is possible to get all the other thermodynamic quantities, internal energy,
specific heat and so on in the usual ways.

It is useful to think about replicated free-energy as the free energy of a
molecular liquid, whose
molecules are composed of $m$ atoms, each in a different replica of the
liquid.  The tendency to
form molecules is forced by a small but extensive coupling term between
particles of different
replicas \cite{Me,Mo}.

In the case of binary mixtures, one may ask if these molecules can be formed
by particles of
different kind.  For the soft sphere binary mixture model \cite{sferesoft} 
the point 
is quite subtle because the particles (and interactions between them) are
different only because of
their effective size, and the ratio of their radii is not so different from 1.
 In the Lennard-Jones
binary mixture model the situation is more clear.  Beside the sizes of
particles, there are three
different values of the potential well for the three different kinds of
interaction $(++,--,+-)$. In this case,
the energy depends much more strongly on the type of particles involved.
In fact, we expect
that even the substitution of a single particle of a given type with a
particle of different type,
at low temperature, substantially increases the total energy of the system,
lowering the probability
of such a configuration which can then be safely neglected.

Since replicated (or
molecular) free energy must describe the free energy of $m$ systems in the
same state, we introduce
attractive coupling terms only between particles of the same kind, neglecting
molecules formed by
atoms of different types.   The replicated partition function is then:

\begin{eqnarray}
\label{zm}
Z_m[\w] & = & \frac{1}{N_{+}^m!N_{-}^m!}
\sum_{\si_a} \sum_{\pi_a} \int \prod_{a}
d^d x_i^a
\exp
\left ( - \frac{\beta}{2} \sum_{i \neq j,a} V ^{\eps_i \eps_j} (x_i^a - x_j^a)
  \right. \nonumber  \\
& - & \left. \sum_{i \in \{+\}} \sum_{a \neq b}
{\w}_+(x_{\si_a (i)}-x_{\si_b (i)}) -
\sum_{i \in \{-\}} \sum_{a \neq b}
{\w}_-(x_{\pi_a (i)}-x_{\pi_b (i)})
\right ),
\end{eqnarray}

\noindent where $\w$ is the attractive potential among replicas and we must
sum over the permutation
($\si_{a}$ and $\pi_{a}$) of the particles in each replica.  When relabeling
particles, so that to
particle $i$ of a given kind in replica $a$ corresponds particle $i$ of the
same kind in replica $b$
(which is supposed to belong to the same molecules) and so on, the sum over
permutations can be
eliminated leaving  a multiplicative factor
$(N^+!  \: N^-!)^{(m-1)}$.

The phase transition we are investigating is equivalent to that
shown by generalized spin glasses \cite{KiThWo}.
The main idea is that in the glass phase there are only few states,
not linked by trivial symmetry transformations,
which contribute to integral (\ref{parti}).
Below $\tk$ these states are mutually inaccessible because of ergodicity
breaking.
In replica space, this spontaneous
one-step replica symmetry breaking is signaled by
the onset of an off-diagonal non trivial correlation,
when couplings are sent to zero.

The study of the thermodynamic phase, where replicas are correlated, is
accomplished by choosing
the probability distribution (after selecting a permutation) of the
$m$ particles inside
the same molecule for both kinds of particles as order parameters:

\beq
\rho_+(r^1,...,r^m)= \sum_{i \in \{+\}} <\delta(x_i^1-r^1)...\delta(x_i^m-r^m)>\\
\label{ord}
\rho_-(r^1,...,r^m)= \sum_{i \in \{-\}} <\delta(x_i^1-r^1)...\delta(x_i^m-r^m)>
\end{eqnarray}

\noindent
where a permutation is already selected, and
introducing the Legendre
trasformate of molecular (replicated) free energy:

\bq
\label{gdia}
G_{m}[\rho] & = & \mathop{\rm lim}_
{
\begin{array}{c}
{\small \w \to 0} \\
{\small N \to \infty}
\end{array}
}
\left(-{1 \over \beta m} \log Z_{m}[\w] - {1 \over m}
\int \prod_{a=1}^m d^d r^a \sum_{\eps = +,-} \rho_{\eps}(r^1,...,r^m)
W_{\eps}(r^1,...,r^m)\right)
\eq
with
\begin{equation}
W_{\eps}(r^1,...,r^m) = \sum_{a < b} \w_{\eps}(r^a-r^b)
\end{equation}

\noindent
Performing the limit $\w \to 0$ is equivalent
to search for a non-trivial saddle point, with respect to the order
parameters, of the functional $G_{m}[\rho]$.
In the presence of a glass transition we expect the
order parameters and thermodynamic quantities to show the following behaviours:

\begin{itemize}
\item For $T > \tk$ the density free energy is the liquid one ($m=1$)
and the order
parameters are trivial, i.e. $\rho_+=c_+ \rho$ and $\rho_-=c_- \rho$
(where the $c$'s are the different concentrations).

\item For $T < \tk$, the density free energy of glass is the maximum of
function $f_m \equiv \Phi(m,T)/m$, that is found at $m^* <1$.
From $f_{m^*}$ we can compute all the thermodynamic quantities.

The free energy and its first derivatives are continuous at $T_K$, while the
heat capacity falls suddenly to solid-like values.
The transition, then, is of second order from thermodynamics point of
view, but it is discontinuous in the order parameter which
abruptly becomes a non trivial function of positions in different replicas.

\end{itemize}

It is natural to describe the particle positions in term of center of
mass coordinates $r_i$ and relative displacements $u_i^a$ with
$x_i^a = z_i + u_i^a$ and $\sum_a u_i^a =0$.
A useful simplification is the choice of quadratic coupling among replicas,
that allows rewriting (\ref{zm}) as:

\bq
Z_m & = & \frac{1}{N^+! \: N^-!} \int \left ( \prod_{i=1}^N d^d z_i \right )
\left (\prod_{a=1}^m \: \prod_{i=1}^N d^d u^a_i \right )
 \left [ \prod_{i=1}^N \left ( m^d \delta ( \sum_{a=1}^m u^a_i) \right )
\right ]  \nonumber \\
 & \cdot & \exp  \left ( -\beta \sum_{a=1}^m
\sum_{i < j } V^{\eps_i \eps_j}(z_i-z_j+ u^a_i-u^a_j) \right.
 \nn \\
& - & \left. \frac{1}{4 \app} \sum_{a, b } \sum_{i \in +} (u^a_i - u^b_i)^2
-\frac{1}{4 \amm} \sum_{a, b } \sum_{i \in -} (u^a_i - u^b_i)^2
\right ),
\label{zmquad}
\eq
where the $\{ u^a_{i\mu} \}$ for a given $i$ (the indices $\mu$ and
$\nu$, running from $1$ to $d$, denote space directions) turn out to be Gaussian
random variables with a vanishing first moment and a second moment given by
\begin{equation}
\langle u^a_{i\mu} \: u^b_{i \nu} \rangle =
\left ( \delta^{ab}-\frac{1}{m} \right ) \delta_{\mu \nu} \delta_{ij}
\frac{\al_{\ei}}{m}.
\end{equation}

\end{section}

\begin{section}{Replicated free-energy}

Three different ways to
compute the 'molecular' free energy of a replicated system were introduced
\cite{MePa2}.
In a previous paper \cite{sferesoft} on binary glasses, free energy was
computed by two of these schemes of approximation, namely
harmonic resummation and small cage approximation; both
gave similar results.
Here we want to unify these two approximations in a single framework.

In the glass phase molecules
are expected to have a small radius, so our starting point is
a quadratic expansion of $V$ in the partition function (\ref{zmquad}).
The integration over these quadratic fluctuations gives:
\bq
Z_m= {m^{Nd/2} \sqrt{2 \pi}^{N d (m-1)} \over {N^+! \: N^-!}}
\int \prod_{i=1}^N d^d z_i \exp\((-\beta m \sum_{i<j}
V^{\eps_i \eps_j}(z_i-z_j) -{m-1 \over 2} Tr \log \((\beta M  \)) \))
\eq
where
the matrix $M$, of dimension $Nd \times Nd$, is given by:
\begin{equation}
\label{matri}
M^{\ei \ej}_{(i \mu) (j \nu)}= \delta_{ij}
\left( \sum_k V^{\ei \ek}_{\mu\nu}(z_i-z_k) + \frac{m}{\al_{\ei}} \right)
-V^{\ei \ej}_{\mu\nu}(z_i-z_j)
\end{equation}
and $V_{\mu\nu}(r) =\partial^2 V /\partial r_\mu \partial r_\nu$.
We have then an effective Hamiltonian where the centers of masses $z_i$ of the
molecules interact at effective temperature $T^*=1/(\beta m)$ by
means of a  pair potential, complicated by the contribution of
vibration modes.
We proceed by using a 'quenched approximation', i.e.
neglecting the feedback of
vibration modes onto the centers of masses:

\bq
Z_m & = & m^{Nd/2} \sqrt{2 \pi}^{N d (m-1)} Z_{liq}(\bb \: m)
\left \langle \exp{\left( -\frac{m-1}{2} \Tr \log \bb M(z) \right) } \right
\rangle_{\bb \: m} \nn \\
& \sim &
m^{Nd/2} \sqrt{2 \pi}^{N d (m-1)} Z_{liq}(\bb \: m)
\exp \left( - \frac{m-1}{2} \left \langle \Tr \log \bb M(z) \right
\rangle_{\bb \: m} \right).
\label{quenched}
\eq
This approximation becomes exact when close to the Kauzmann temperature
where $m \to 1$.

We normalize matrix elements as follows:

\begin{equation}
\label{ele}
C^{\eps \epp}_{(i\mu)(j\nu)} \equiv
\sqrt{c_{\eps} c_{\epp} \over r_{\eps}^2} V^{\eps \epp}(z_i-z_j).
\end{equation}
introducing the mean values of diagonal terms:

\begin{equation}
\label{norma}
r_{\eps} =  \sum_{\epp} c_{\eps} \rho \int{d^d r g_{\eps \epp}(r) \frac{1}{d}
\Delta V^{\eps \epp}(r) } + \frac{m}{\al_{\eps}}.
\end{equation}

\noindent
The replicated free energy is:

\bq
\label{replicata}
{\beta \phi(m,\beta)} & = &
-{d \over 2 m} \log(m)- { d (m-1) \over 2 m } \log(2 \pi)
-{1 \over m N} \log Z_{liq}(\beta \: m) \nonumber \\
& + & \frac{d \:(m-1)}{2m} \left (c_+ \: \log(\beta \: r_+)
+c_- \log(\beta \: r_-) \right )  \nn \\
& + & {1 \over N} {(m-1) \over 2 m} \left \langle
\Tr \log\left[ \delta_{ij} \sum_k
C^{\eps \epp}_{(i\mu)(k \nu)}  -
C^{\eps \epp}_{(i\mu)(j \nu)} \right] \right \rangle.
\eq
We consider fluctuations of diagonal term $\sum_k C_{ik}$ up to second
order, while the whole non-diagonal contribution $C_{ij}$ is resummed
by means of a kind of chain approximation.
The free-energy to compute is, then:

\bq
\label{fharmoin}
{\beta \phi(m,\beta)} & = &
-{d \over 2 m} \log(m)- { d (m-1) \over 2 m } \log(2 \pi)
-{1 \over m N} \log Z_{liq}(\beta \: m)  \nonumber \\
& + & \frac{d \:(m-1)}{2m} \left (c_+ \: \log(\beta \: r_+)
+c_- \log(\beta \: r_-) \right )  \nn \\
& - & {1 \over 2 N} {(m-1) \over 2 m} \left( \langle
\Tr C^{\eps \epp}_{(i\mu)(k\nu)}
C^{\eps \epp}_{(i\mu)(k^{\prime}\nu)} \rangle -dN \right) \nn \\
& + & \frac{1}{N} {(m-1) \over 2 m}  \sum_{p=2}^{\infty}
 \left \langle \frac{\Tr C^p}{p}  \right \rangle.
\eq

\noindent
The $p$-th order term depends, as usual, on the $p$-point correlation
function

\bq
\left \langle Tr C^p \right \rangle & = & \sum_{\eps_1...\eps_p \in \{+,-\}}
\sum_{\mu_1...\mu_p}
 \int d^d z_1 \dots d^d z_p \: \rho^p
g^{\eps_1 \dots \eps_p}(z_1 \dots z_p) C^{\eps_1\eps_2}_{\mu_1\mu_2}(z_1-z_2)
 \nn \\
  & \cdots & C^{\eps_{p-1}\eps_p}_{\mu_{p-1}\mu_p}(z_{p-1}-z_p)
  C^{\eps_p \eps_1}_{\mu_p\mu_1}(z_p-z_1),
\eq

\noindent
The 'chain' approximation in the computation of the traces
neglects contributions coming from the terms where two indices
coincide and utilizes the
superposition approximation to obtain $p$-point correlation
functions $g^{(p)}(z_1...z_p)=g(z_1-z_2) \cdots g(z_p-z_1)$.
With these hypotheses we arrive at:

\bq
\left \langle \Tr C^p \right \rangle & = &
\int d^d z_1 \dots d^d z_p \: \rho^p \:  \sum_{\mu_1 \dots \mu_p}
\sum_{\eps_1...\eps_p}
g^{\eps_1\eps_2}(z_1-z_2)   C^{\eps_1\eps_2}_{\mu_1 \mu_2}(z_1-z_2)
 \nn \\
& \cdots & g^{\eps_{p-1}\eps_p}(z_{p-1}-z_p ) C^{\eps_{p-1}\eps_p}_{\mu_{p-1}
\mu_p}(z_{p-1}-z_p )
g^{\eps_p\eps_1}(z_p-z_1) C^{\eps_p\eps_1}_{\mu_p \mu_1}(z_p-z_1).
\eq

\noindent
Traces are computed in Fourier space, where the matrix is diagonal
with respect to particle indices, by introducing

\begin{equation}
D_{\mu\nu}^{\eps \epp} (k) \equiv
\int d^d r
g^{\eps \epp} (r) C^{\eps \epp}_{\mu\nu}(r) e^{i k r},
\label{defab}
\end{equation}

\noindent
which can be decomposed in its diagonal (longitudinal) and
traceless (transversal) parts with respect to spatial indices

\begin{equation}
D_{\mu\nu}^{\eps \epp} (k)=\delta_{\mu\nu} \ a^{\eps\epp}(k) +
\(( {k_\mu k_\nu \over k^2} -{ \delta_{\mu \nu}\over d} \))
b^{\eps \epp}(k).
\end{equation}

\noindent
The last step consists in the diagonalization of $D$ in the space of
components.
For each $k$, one has four  distinct eigenvalues, the two 'longitudinal' ones,
corresponding to those of

\begin{equation}
D^{\eps\epp}_{\parallel}(k)=a^{\eps \epp}(k)+ {d-1 \over d} b^{\eps \epp}(k),
\end{equation}

\noindent
and the two 'transverse' eigenvalues of the matrix

\begin{equation}
D^{\eps \epp}_{\perp}(k)=a^{\eps \epp}(k)-{1 \over d} b^{\eps \epp}(k).
\end{equation}

\noindent
The eigenvalues are:

\bq
\lambda_\parallel&=&{1 \over 2} \((D^{++}_{\parallel}+D^{--}_{\parallel} + \sqrt{
(D^{++}_{\parallel}-D^{--}_{\parallel})^2+4 (D^{+-}_{\parallel})^2} \))
\nn \\
\mu_\parallel&=&{1 \over 2} \((D^{++}_{\parallel}+D^{--}_{\parallel} - \sqrt{
(D^{++}_{\parallel}-D^{--}_{\parallel})^2+4 (D^{+-}_{\parallel})^2}\))
\nn \\
\lambda_\perp&=&{1 \over 2} \((D^{++}_{\perp}+D^{--}_{\perp} + \sqrt{
(D^{++}_{\perp}-D^{--}_{\perp})^2+4 (D^{+-}_{\perp})^2} \))
\nn \\
\mu_\perp&=&{1 \over 2} \((D^{++}_{\perp}+D^{--}_{\perp} - \sqrt{
(D^{++}_{\perp}-D^{--}_{\perp})^2+4 (D^{+-}_{\perp})^2}\)).
\eq

\n
The higher order terms involving the fluctuation of $\sum_k C_{ik}$
cannot be resummed in a systematic way similar to the one
introduced for the off-diagonal terms $C_{ij}$.
In any event, neglecting them is not likely to dramatically change the final result,
because the sum of all terms higher than order two contributes just a
small correction to second order computation for non-diagonal matrix elements.
This is confirmed by comparison to simulations.

In addition, the second order term from these calculations is equal to the second order term
computed in the small cage approximation when non-harmonic corrections
due to higher derivative are neglected.
Summarizing:

\begin{itemize}
\item
The harmonic resummation scheme 
\cite{MePa1,MePa2,sferesoft} allows us to resum the whole contribution coming from
off-diagonal matrix elements, but it misses contributions from non
harmonic corrections and from thermal fluctuation of diagonal elements
of $M$.

\item
Both kinds of correction are accounted by
the small cage approximation
\cite{MePa2,sferesoft}, but in that case we are unable to
systematically resum either the terms of the series:
\begin{equation}
\label{serie}
\sum_{p=2}^{\infty} \left \langle \frac{\Tr C^p}{p}  \right \rangle.
\end{equation}
or the contributions from non-harmonic corrections, and one has to
truncate the expansion at some order (up to now we truncated at the 
second order).

\item
It is possible to partially merge second order contributions
from the small cage
computation in the harmonic resummation scheme, obtaining the whole
resummed series plus fluctuations of
diagonal elements up to second order.

\end{itemize}

The final expression we get for the binary mixture free energy within
this scheme is:

\bq
\label{fharmofin}
\phi(m,\beta) & = & -{d \over 2 m} \log(m)- { d (m-1) \over 2 m } \log(2 \pi)
 + {d (m-1) \over 2 m } \((c_+\log (\beta r_+) + c_-\log(\beta r_-) \))
 \nn \\
&-& {(m-1) \over 2 m } \frac{1}{\rho}
\int \frac{d^d k}{2 \pi^3} \sum_{\eps \epp \epd}
d \: a^{\eps \epp}(k)  a^{\epp \epd}(k)  h^{\epd \eps}(k)
+ \left( 1 - {1 \over d} \right)
b^{\eps \epp}(k)  b^{\epp \epd}(k)  h^{\epd \eps}(k) \nn \\
& + &
{(m-1) \over 2 m } \frac{1}{\rho}
\int  \frac{d^d k}{2 \pi^3}  \left\{ L_3 (\lambda_\parallel(k)) + L_3 (\mu_\parallel(k))
+
{(d-1)} \left[ L_3 (\lambda_\perp(k)) + L_3 (\mu_\perp(k))\right] \right\}
 \nn \\
&-& {(m-1) \over 2 m } \int d^d r \: \rho \sum_{\eps \epp}
g^{\eps \epp}(r) \sum_{\mu\nu} \((C^{\eps \epp}_{\mu\nu}(r)\))^2
-{1 \over m N} \log Z_{liq}(\beta \: m),
\label{chain}
\eq

\noindent
where the function $L_3$ is $\log(1-x)+x+{x^2/2}$.
Let us point out that the term in the second line is just the contribution,
to lowest order, due to fluctuations of the diagonal elements of the Hessian
matrix.

It is now simple to compute the equilibrium  complexity $\Sigma_{eq}$:

\begin{equation}
\Sigma_{eq}(\beta) =
m^2 {\partial \beta F_m \over \partial m }{\Biggr |}_{m=1}=S_{liq}-S_{sol},
\label{comple}
\end{equation}
where
$S_{liq}$ is the entropy of the liquid at the effective temperature
$T_{eff}$, which equals $T$ for $m=1$, and  $S_{sol}$ is the entropy of a
harmonic solid with a matrix of second derivatives given by $M$, i.e:

\begin{equation}
S_{sol}={d \over 2} \log(2 \pi) - {1\over {2 \: N}}\la Tr \log\((\beta M  \))\ra.
\end{equation}

The condition
for identifying the Kauzmann temperature, $\Sigma_{eq}(\bb)=0$,
reads in our approximation simply:

\begin{equation}
S_{liq}=S_{sol},
\end{equation}
as expected on general grounds \cite{Mo}.

If from our computation we find that $S_{liq}<S_{sol}$, we are already  in the
glass phase
($T<T_K$) and the computation in the liquid does not make sense,
while in the other case $S_{liq}>S_{sol}$, the temperature is greater than
$T_K$ (and of course less
than $T_D$ if the spectrum of $M$ is positive).

With (\ref{fharmofin}) we have a tool to investigate the thermodynamics of
the low temperature glass phase. It should be emphasized that
only the liquid phase $g(r)$ and free-energy are needed in order to compute
the glass phase thermodynamics.
Beyond the usual thermodynamic quantities
(energy, entropy, heat capacity...), we are interested in the two new
parameters describing the glass phase:

\begin{itemize}

\item
The square cage radii $A_{\eps}$, defined as $A_{\eps}= {1 \over 3} (
\la x_i^2\ra -\la x_i \ra^2)$ for type $\eps$ particles.
These square cage radii are obtained by differentiating the free energy
with respect to coupling terms and by sending couplings to zero at the end:

\begin{equation}
A_{\eps} = {2 \over d (m-1) N_{\eps}} \  {\partial (\beta F) \over
\partial (1/ \aeps)} (\aeps=\infty).
\label{formugabbie}
\end{equation}

\n
This expression gives square cage radii nearly linear in
temperature in the whole glass  phase (see figure [6b]),
which is natural since non-harmonic effects have been neglected.
The values of $A_{\eps}$ at the Kauzmann temperature are of the same
order of those obtained for the soft sphere model, i.e. $ \sim 10^{-3}$.

\item
The effective temperature $\teff=T/ m^*$ of the molecular liquid.
This varies very little and remains close to
the Kauzmann temperature throughout the low temperature phase, confirming the
hypothesis that glass can be successfully described by means of a system
of molecules remaining in the liquid phase.
We stress that the linear behaviour of parameter $m^*$ as a function
of $T$ is a feature shared by every 1RSB system to our knowledge.
This static parameter should be equivalent to the dynamical parameter that
 measures
violation of the fluctuation-dissipation theorem \cite{FrMePaPe}, as it is observed in
short range spin-glasses \cite{MaPaRiRu2}

\end{itemize}

The harmonic expansion makes sense only if  $M$ has no negative eigenvalues.
This is natural since it is intimately related to the vibration modes of the glass.
 Notice that we cannot describe activated processes here,
and so we cannot see the tail of negative eigenvalues (with number
decreasing as
$\exp(-1/T)$ at low temperatures), which is always present. It is known however
that the fraction of  negative eigenvalues of $M$ becomes negligible below
 the dynamical transition temperature, $T_D$ \cite{ScTa,Ke}.
Therefore our harmonic expansion makes sense if the effective temperature, $\teff$,
is less than $T_D$.

\end{section}

\begin{section}{Integral equations for binary mixtures}

Unlike simpler models such as the soft sphere model \cite{sferesoft},
the glass transition for the considered Lennard-Jones binary mixture cannot be
described in the simple framework of well-known approximations of
nonlinear integral equations for liquids: the Hypernetted Chain
approximation (HNC); the
Mean Spherical Approximation (MSA); the Percus-Yevick Approximation (PY)
\cite{Han1}.
Within all the classical approximation HNC, PY and MSA, 
the compressibility, computed as an integral of correlation functions:

\begin{equation}
\label{intcompre}
\rho \chi_1 = \bb + \rho \bb \int{d^3 r h(r)},  \nonumber
\end{equation}
diverges at a finite temperature $T_{GL}$ which depends
on the chosen approximations [fig. 1].
We draw two conclusions from these computations.
Our first conclusion is that this singularity
corresponds to the liquid-gas phase separation which is typical
of simple systems \cite{Han1}.
To convince the reader, in [fig. 2] we show the
phase diagram in the plane $\rho-T$ computed within HNC approximation,
where the region of phase coexistence is shown. 

Hence, our claim is that HNC and PY, in this model,
badly overestimate the real liquid-gas phase
separation temperature, $T_{GL}$, while MSA approximation
allows to get far closer to this transition.
This is confirmed by our
numerical simulations (see below) placing $T_{GL}$ for
$\rho = 1.2$ at $ \sim 0.3$.

\begin{figure}[htbp]
\begin{center}
\leavevmode
\epsfig{figure=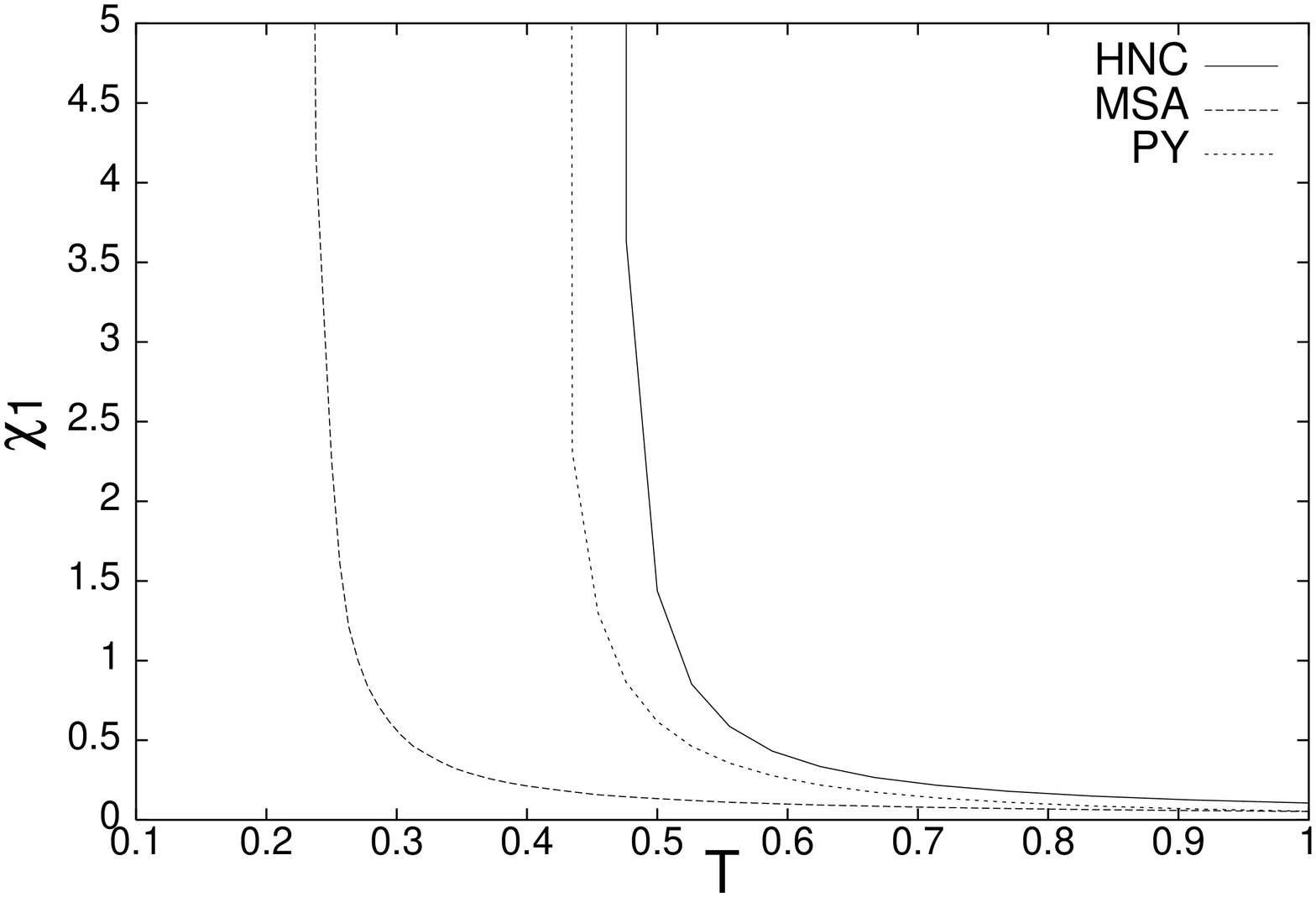,angle=270,width=11cm}
\caption{Compressibility, $\chi$, as a function of temperature, $T$,
computed by HNC (filled line), PY (dashed line) and MSA (dotted line)
approximations.}
\end{center}
\label{compreHPM}
\end{figure}

\begin{figure}[htbp]
\begin{center}
\leavevmode
\epsfig{figure=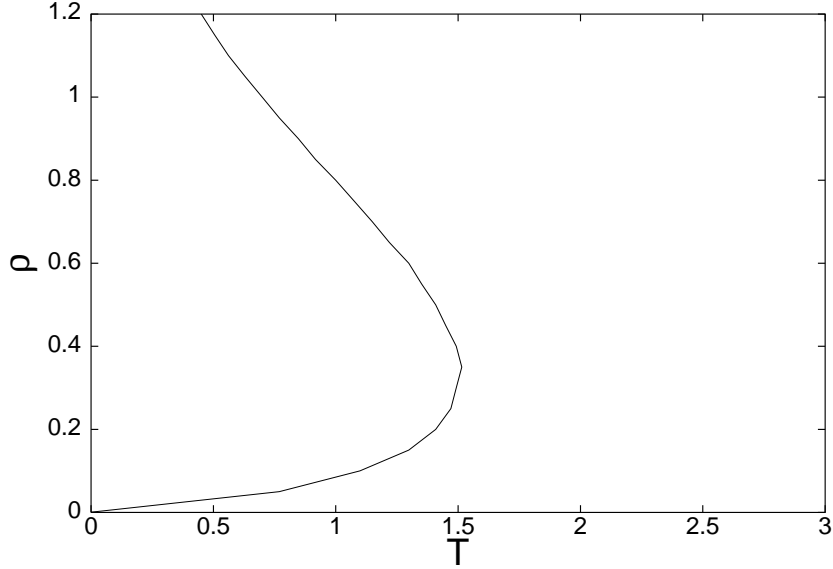,angle=270,width=11cm}
\caption{Projection of the phase diagram in the $\rho-T$ plane
computed by HNC approximation. In the region on the left of the line,
a thermodynamically stable homogeneous phase does not exist and the
system undergoes a liquid-gas phase separation.}
\end{center}
\label{phasedia}
\end{figure}

A simple procedure for improving the integral equations has been proposed
by Zerah and Hansen \cite{ZeHa}. It mixes the HNC and MSA closures
by means of a single parameter, $\al$, that is chosen in order to
reduce thermodynamic inconsistencies between the two different
routes for computing the compressibility:

\bq
\label{compre}
\rho \chi_1 & = & \bb + \rho \bb \int{d^3 r h(r)}  \nonumber \\
\rho \chi_2 & = & \left( \frac{d P}{d \rho} \right)^{-1}
\eq
where the pressure, $P$, is computed via the virial equation:

\begin{equation}
\label{pressione}
\frac{\bb P}{\rho} = 1 - \frac{2}{3} \pi \beta \rho \int{d r \: r^3 \:
  V^{\prime}(r)
g(r)}
\end{equation}

Briefly, we recall that
HNC approximation consists of neglecting the 'bridge' diagrams in
the cluster Mayer expansion, while MSA
treats the attractive term of the potential as a correction to the repulsive
term.

The 'mixed' integral equation which we have resort to is, then:

\bq
\label{hmsa}
c^{\eps} c^{\epp} \rho^2 g_{\eps \epp}(r) & = & \exp
\left( - \beta V^{\eps \epp}_{(R)}(r) \right)
\left( 1+ \frac{\exp
\{ f^{\eps \epp}(r) [ w^{\eps \epp}(r) - \bb  V^{\eps \epp}_{(A)}(r)
] \} -1}{f^{\eps \epp}(r)} \right)
\eq
where $V_{(A)}(r)$ and $V_{(R)}(r)$ are, respectively, attractive and
repulsive contribution to potential:

\bq
V_{(R)} & \equiv & \left \{
\begin{array}{lcl}
V(r) - V_{min} & \hspace{.5cm} & r \leq r_{min} \\
  0 & \hspace{.5cm}  & r \geq r_{min} \end{array} \right. \nonumber \\
V_{(A)} & \equiv & \left\{
\begin{array}{lcl}
V_{min} & \hspace{2cm} &  r \leq r_{min} \\
V(r)  & \hspace{2cm} &  r \geq r_{min} \end{array} \right.
\eq
and the function $W(r)$ is $h(r)-c(r)$, with $c(r)$ the direct
correlation function.

In the Zerah and Hansen scheme
the function $f^{\eps \epp}(r)$ is introduced,
allowing continuously interpolation between HNC and MSA, simply by choosing:

\begin{equation}
\label{interpola}
f^{\eps \epp}(r) = 1 - \exp \left( - \frac{r}{\s_{\eps \epp} \al} \right).
\end{equation}
Let us stress that for $f(r) \to 1$ the closure (\ref{hmsa}) reduce to
HNC, while in the opposite limit, $f(r) \to 0$, MSA is recovered.
The choice of (\ref{interpola}) implies that we obtain correlation
functions which are HNC-like at large $r$ and MSA-like at short $r$.

We have chosen to utilize this scheme in a different
fashion, since we are mainly interested to obtain  
pair correlation functions which very accurately reproduce the
observed liquid behaviour in order to study the effects of the 
used approximations in describing the liquid-glass transition. 
The HNC closure is exact at very high temperatures and,
within this model, 
significant deviations from it of the numerical data of the energy appear
yet in the region $\bb \in [10^{-4},0.1]$.
On the other hand, the MSA approximation turns out to be more correct
when the liquid-gas coexistence region is approached, both in 
reproducing the numerical data and in reducing the inconsistencies 
between the two different compressibilities.
We have chosen as a starting value $\alpha=1.5$ at $\bb=0.1$,
which fits very well the numerical data in 
the whole region $\bb \in [10^{-4},0.1]$ and we increased it linearly
in such a way to have a MSA value ($\al=5$) at the liquid-gas
transition.

The result is a 
very good agreement between internal energy computed with the  
$g(r)$ obtained by solving (\ref{hmsa})
and the results of numerical simulations [fig.3]. 
These will be discussed in detail
in the following. Let us remind that the pure HNC and MSA closure 
have an error up to 10\%, 
while, with interpolated integral equations, it is greatly reduced.

\begin{figure}[htbp]
\begin{center}
\leavevmode
\epsfig{figure=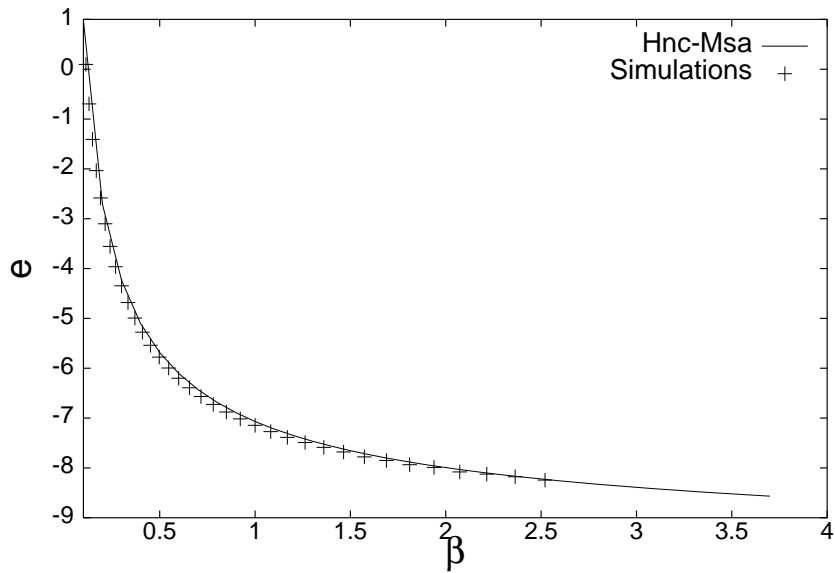,angle=270,width=11cm}
\caption{Internal energy vs. inverse temperature, $\beta$.
The points are obtained by simulated annealing runs,
the solid line by interpolated integral equation}
\end{center}
\label{enesim}
\end{figure}

\end{section}

\begin{section}{Results and discussion}

We have studied a binary mixture where particles interact by means of a
Lennard-Jones type potential. This model has been introduced and
extensively studied as a good glass former.
For density $\rho=1.2$, numerical simulations show evidence for a
'dynamical' transition at temperature $T_{D} = 0.435$ \cite{KoAn}.
This kind of transition does not correspond to a real thermodynamic
transition with singularities in thermodynamic quantities, but rather seems
related to changes in the relaxation processes of
the system in the free-energy landscape.
Mean field theory of supercooled liquid, Mode Coupling Theory
\cite{GoSj}, describes this dynamical transition.

Complementary to this 'dynamical' picture of the glass transition
is the investigation of the free-energy of a 'molecular' liquid,
where 'molecules' are formed by $m$ replicas of the particles in the
original system, as recently proposed \cite{MePa1,Me,MePa2} and
studied in this paper for a Lennard-Jones model.

Following this strategy, we show that at the temperature $\tk = 0.32$
(see the figures below) the complexity, i.e. the
entropy due to the large number of minima in potential energy as a
function of particle positions, vanishes.
In our 'thermodynamic' picture, $\tk$ corresponds
to the Kauzmann temperature, where the
supercooled liquid undergoes a true thermodynamic transition and the
system enters a new phase, where it behaves as an amorphous solid.

\begin{subsection}{Thermodynamic quantities}

From the thermodynamic point of view, the transition is  second order.
Indeed, the free energy and the internal energy have no singularity
[fig. 4], while
one can observe that the specific heat [fig. 5]
shows an evident 'jump'
at $T_K$, remaining close to $3/2$, in agreement with the Dulong-Petit law,
throughout the glass phase.

\begin{figure}[htbp]
\begin{center}
\leavevmode
\epsfig{figure=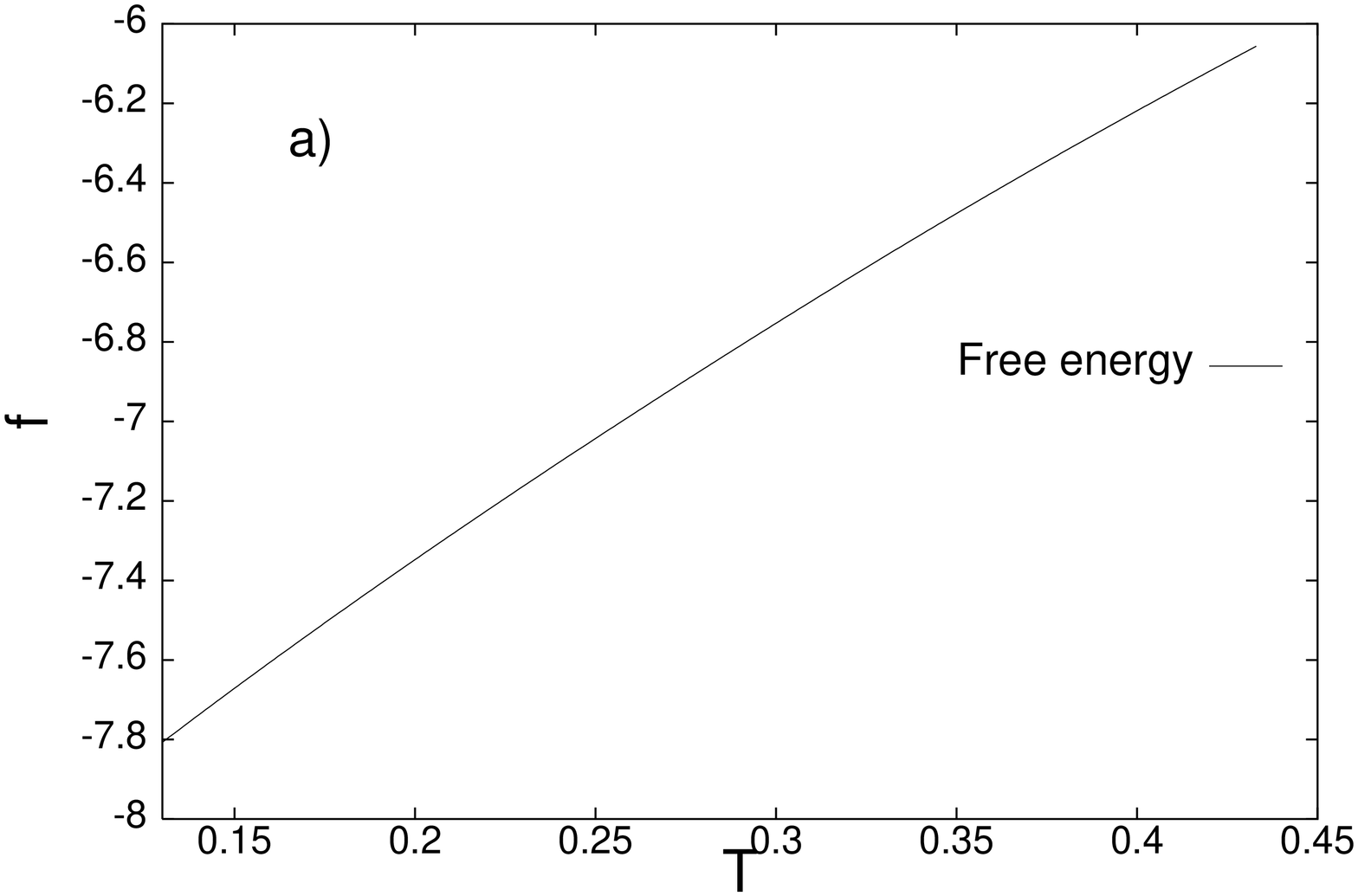,angle=270,width=8cm}
\epsfig{figure=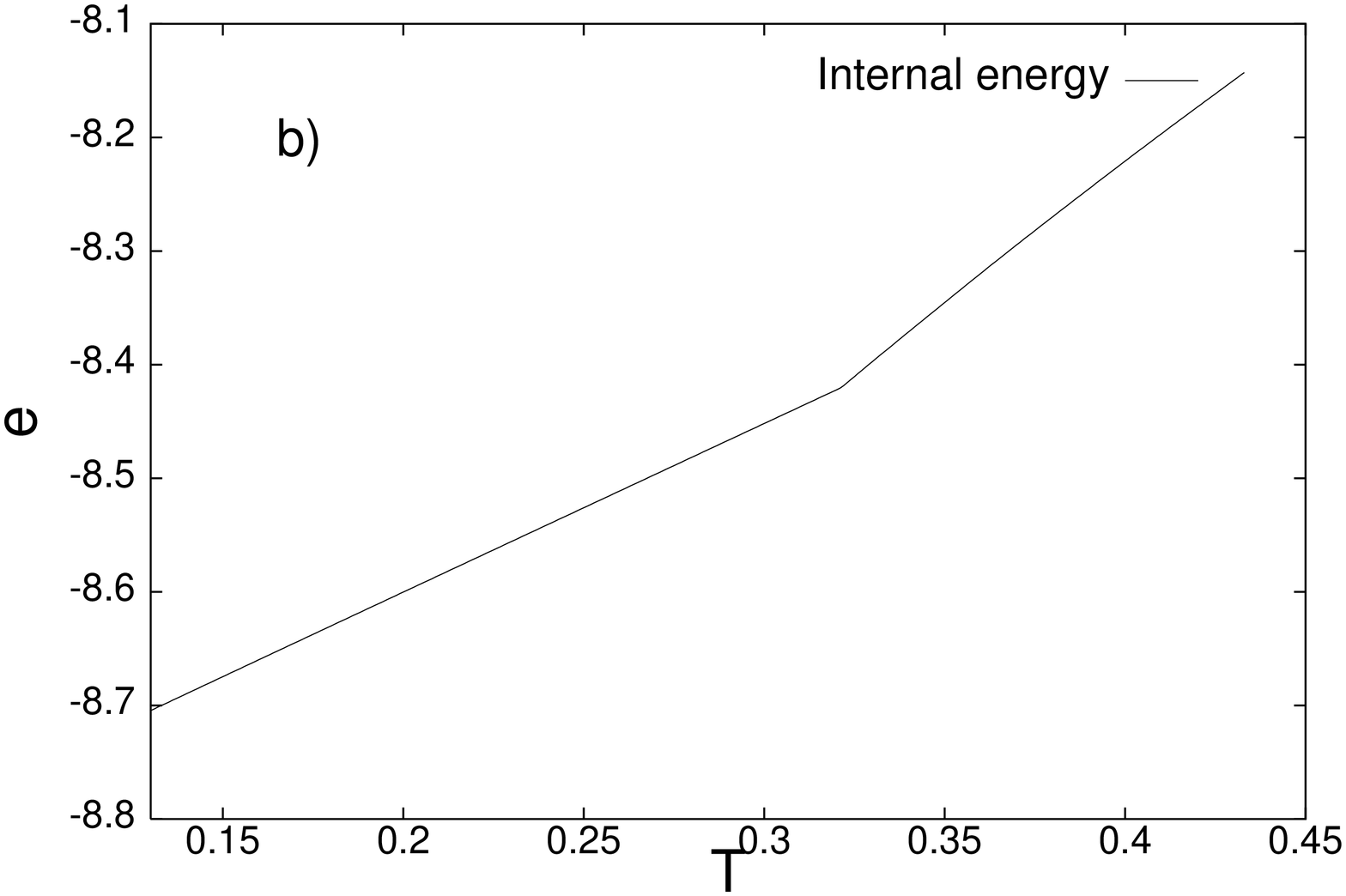,angle=270,width=8cm}
\caption{The free energy (a) and the internal energy  (b)
of the Lennard-Jones model as a function of the temperature,
in both the liquid and the glass phase.}
\end{center}
\label{enes}
\end{figure}

\begin{figure}[htbp]
\begin{center}
\leavevmode
\epsfig{figure=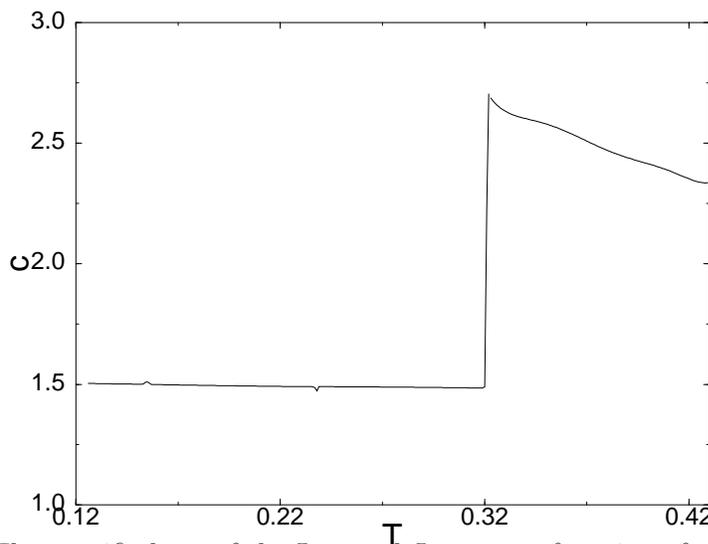,angle=270,width=11cm}
\caption{The specific heat
of the Lennard-Jones as a function of the temperature.}
\end{center}
\label{calspec}
\end{figure}

\end{subsection}

\begin{subsection}{New thermodynamic parameters}

In the framework of replica theory of structural glasses two new
parameters are introduced, analogous to those used to
describe the glass phase in spin-glass models:
the parameter $m$, plotted in [fig. 6a]
and cages size, $A^+$ and $A^-$,  plotted in [fig. 6b].
Both are nearly linear with temperature.
In particular, this means that
the effective temperature, $\beta \: m$, is always close to the
transition value, so in our theoretical computation we need
only the mean values of observables at temperatures $T > T_{GL}$.

\begin{figure}[htbp]
\begin{center}
\leavevmode
\epsfig{figure=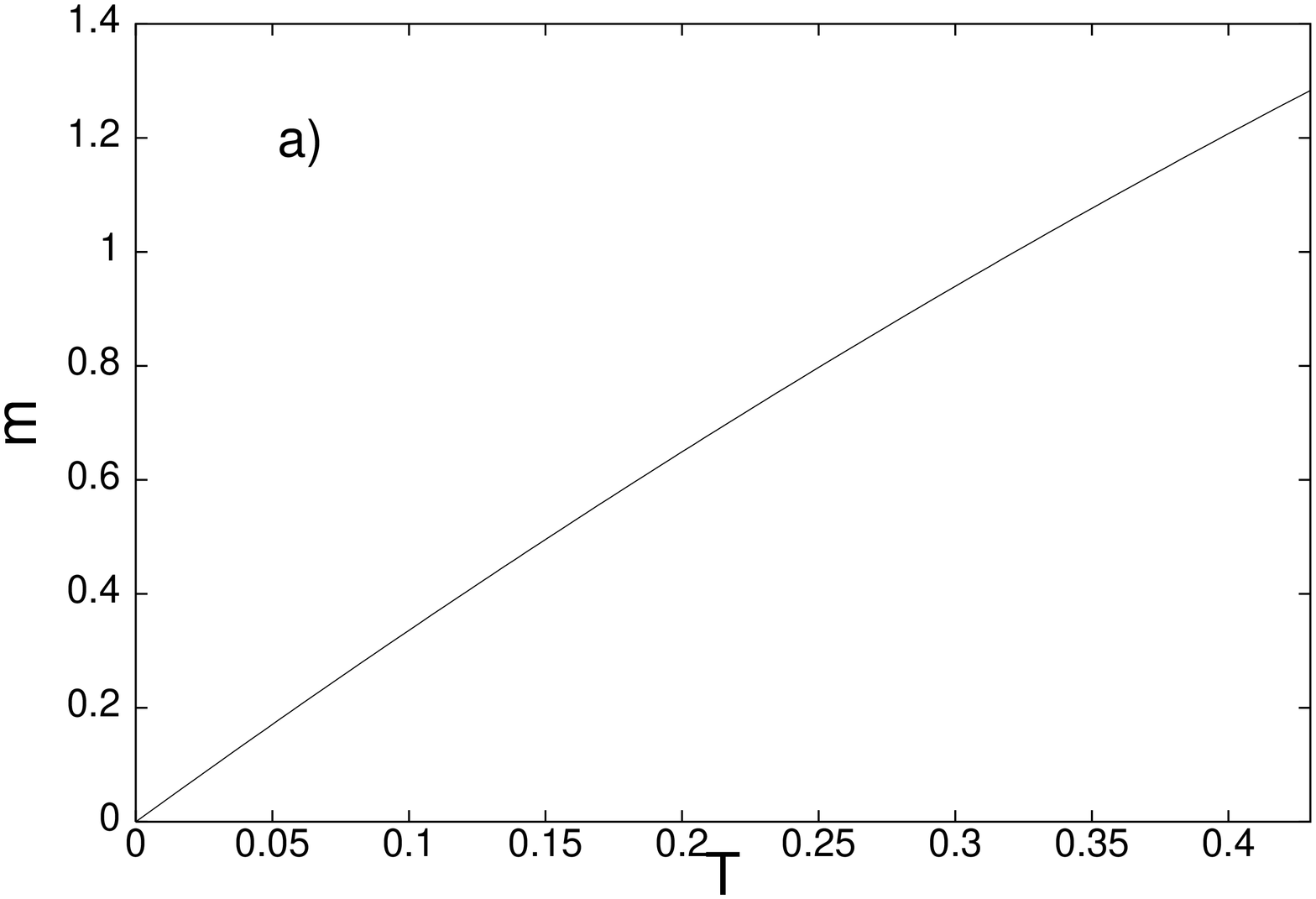,angle=270,width=8cm}
\label{emme}
\epsfig{figure=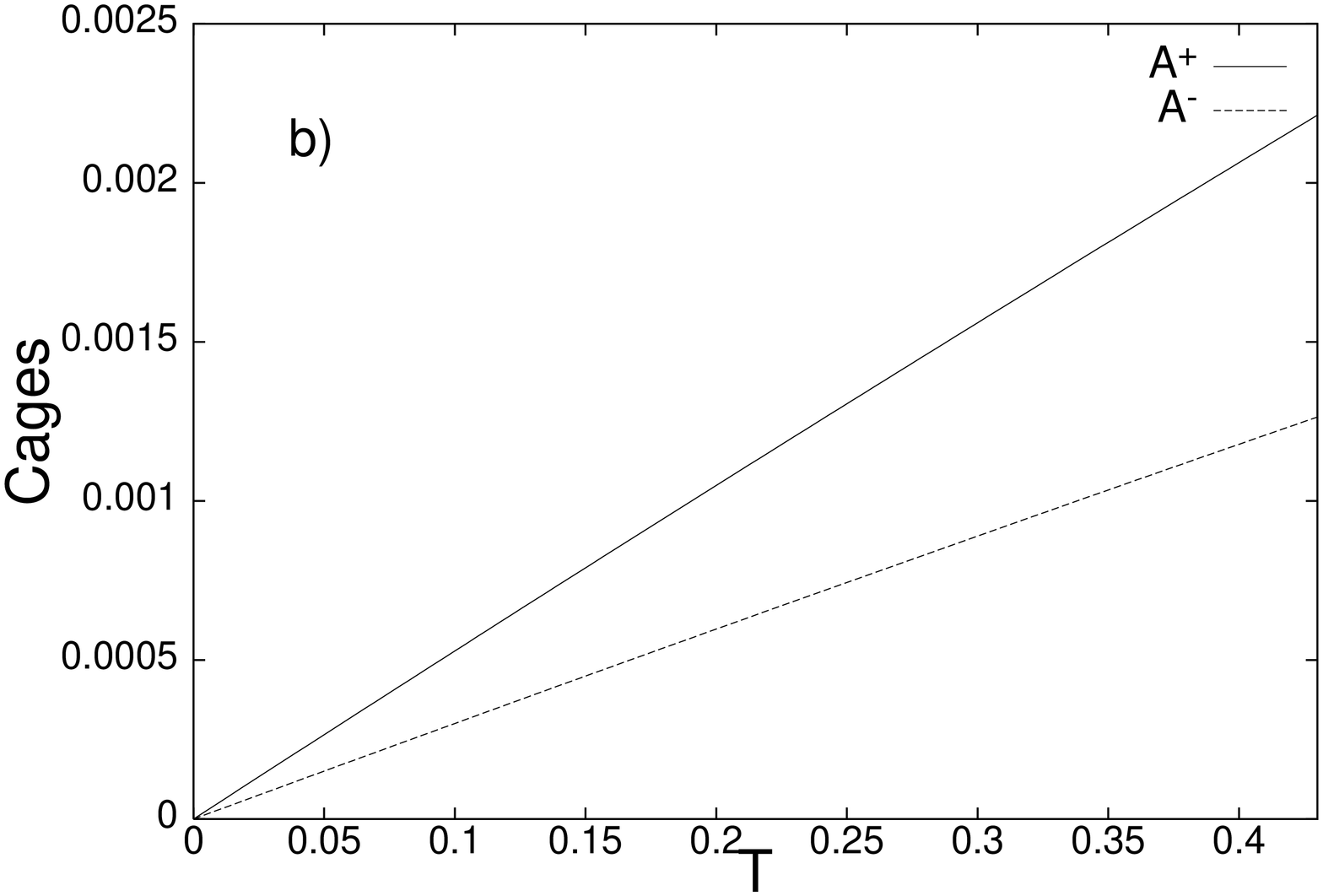,angle=270,width=8cm}
\label{gabbie}
\caption{$m$ (a) and the cage radii $A^+$ and $A^-$ (b) as functions
  of the temperature.}
\end{center}
\end{figure}

\end{subsection}

\begin{subsection}{Comparisons with MC simulations}

Before discussing the numerical results on the complexity and the
evaluation of the transition temperature, let us pay some attention to the
possible presence of a liquid-gas phase separation at a non zero $T_{GL}$.
We identify this temperature as the one where the liquid pressure
$P_{liq}$ equals the gas pressure $P_{gas}$.
At low enough temperatures, the coexistence is between liquid and gas at
very small density, where $P_{gas} \simeq \rho/\beta$ is negligible.
Therefore, as a first approximation, we evaluate  $T_{GL}$ as the temperature
where the liquid pressure becomes compatible with zero.
This pressure is computed via the virial equation:
\begin{equation}
\frac{P}{\rho}=\frac{1}{\beta}-\frac{1}{6 \: N} \left \langle
\sum_{i \neq j} r_{ij}
\frac{d V^{\ei \ej}}{d \: r_{ij}} (r_{ij}) \right \rangle.
\end{equation}
We present in [Fig 7] data on the behaviour of
${1 \over 6} \langle r d V/dr \rangle$, that seems very well fitted
by the power law $a T^{3/5}+b$, usually encountered in liquids
\cite{RoTa}.
By extrapolating data at temperatures lower than the ones where we
succeeded in thermalizing the system, we get a definitely non
zero $T_{GL} \sim 0.3$.

\begin{figure}[htbp]
\begin{center}
\leavevmode
\centerline{\epsfig{figure=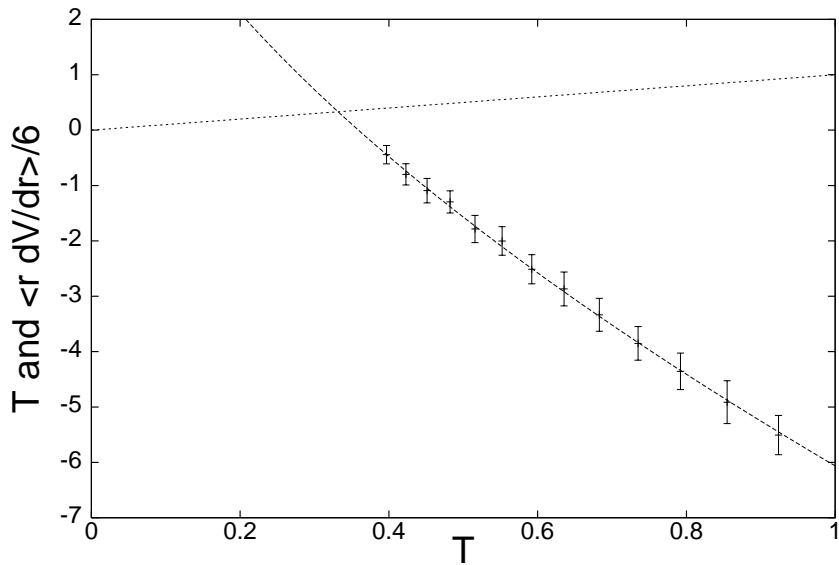,angle=270,width=11cm}}
\caption{The numerical evaluation of $T_{GL}\sim 0.3$ as the
  temperature where the extrapolation of data on ${1 \over 6} \langle r
  d V/dr \rangle$ equals $1/\beta$, i.e. the liquid pressure becomes
compatible with zero.}
\end{center}
\label{complex}
\end{figure}
Though more extensive investigations are needed to confirm this result,
our data are in good agreement with the existence of a liquid-gas phase
separation.
Moreover, they strongly support our hypothesis that approximation
schemes like HNC and PY overestimate the transition temperature.
Therefore, introducing the Zerah and Hansen scheme of interpolation
is necessary to analytically evaluate liquid quantities near the
transition point. Let us recall that this closure gives
$T_{GL} \simeq 0.25$, a slightly lower value than numerical simulations.

Details of the simulations will be discussed below but we would
like to emphasize here that the model we considered is the LJ binary mixture
described in (\ref{model}) without any cutoff on the potentials,
taking into account the small correction due to finite size
effects.
The presence of a cutoff (it is usual to take
$V^{\epsilon \epsilon'}(r)=0$ for
$r > 2.5 \sigma^{++}$) would obviously influence both the energy
and the pressure behaviour and we expect a lower value for $T_{GL}$.

In order to evaluate the complexity and the thermodynamic
temperature by simulations, we consider the numerical approach 
suggested by the harmonic resummation scheme, first utilized for 
the soft spheres binary mixture model \cite{sferesoft}.
We measure both the liquid and the amorphous solid
entropy, whose difference gives $\Sigma$.
$T_K$ is the temperature where
the entropies become equal.

The liquid entropy can be obtained
by numerically integrating the energy density
\begin{equation}
S_{liq}(\beta)=\beta \left ( e_{liq}(\beta)-f_{liq}(\beta) \right )=
S_{liq}^0 + \beta \: e_{liq}(\beta) -\int_0^\beta d \beta' e_{liq}(\beta')
\end{equation}
where $S_{liq}^0$ is the entropy of the perfect gas in the $\beta \rightarrow
\infty$ limit, i.e in the binary mixture case
\begin{equation}
S_{liq}^0= 1-\log \rho-c\log c-(1-c)\log(1-c).
\end{equation}

As it has been discussed in the previous paper \cite{sferesoft}, the 
numerical computation of the
'harmonic solid' entropy is a more subtle task. One has
\begin{equation}
S_{sol}(\beta) = {d \over 2} (1+\log(2 \pi)) - {1\over 2 \: N}
\left \langle  \Tr \log (\beta M ) \right \rangle,
\end{equation}
but the measure of $S_{sol}$ from the 'instantaneous' Hessian is
complicated by the negative eigenvalues,
always present beyond the mean field approximation,
though their number decreases as $\exp(-1/T)$ at low temperatures
and it is expected to be negligible below the Mode Coupling
temperature \cite{ScTa,Ke}.

We consider three different ways of evaluating $S_{sol}$:
\begin{itemize}
\item We measure it from only the ${\cal N}_{pos}$ positive eigenvalues
\begin{equation}
S^{(a)}_{sol} ={d \over 2} \left [ \left (1+\log({2 \pi \over \beta}) \right) -
\left \langle \frac{1}{{\cal N}_{pos}} \sum_{i=1}^{{\cal N}_{pos}} \log
{\lambda_i} \right \rangle \right ].
\end{equation}
\item We consider also the absolute values of
the negative ones (we are disregarding both here and in the previous
case the very few $|\lambda| < 10^{-4}$),
\begin{equation}
S^{(b)}_{sol} ={d \over 2} \left [ \left (1+\log({2 \pi \over \beta}) \right) -
\left \langle \frac{1}{{ d N}} \sum_{i=1}^{d N} \log {|\lambda_i|}
\right \rangle \right ].
\end{equation}
\item Starting from an equilibrium configuration at a given
$\Gamma$ value, we perform a Monte Carlo run at $T=0$, allowing only small
displacements to each particle. The percentage of non-positive eigenvalues
becomes very rapidly
$< 2 \%$  in the whole temperature range considered and, correspondingly,
we get compatible results on $S^{(c)}_{sol}$
when using only the positive eigenvalues
or also the negative ones.
\end{itemize}

As shown in [fig. 8], there is not a small difference between
$S^{(a)}_{sol}$ and $S^{(b)}_{sol}$, particularly at temperatures $T \simg T_D$.
This is also related to the fact that
nearly all the negative eigenvalues are less than one in
absolute value. On the other hand, $S^{(c)}_{sol}$ definitely appear
nearer to the 'instantaneous' Hessian entropy evaluated by taking all the
eigenvalues in absolute value. This has also been observed 
\cite{sferesoft} in the case of the soft sphere binary mixture and it suggests that
$S^{(b)}_{sol}$ is a more reasonable estimation of the solid entropy.

The other problem that we face is the well-known and difficult task of thermalizing
glass-forming liquids at
low temperatures.  We choose to perform a simulated annealing run of a rather
large system, using
data on the liquid energy down to the temperature where the equilibrium was
still reachable in a
reasonable CPU time ($\beta \sim 2.5$).

Then we extrapolate the liquid entropy behaviour at lower temperatures by
fitting data
in the interval $\beta \in [1,2.5]$ with the power law
\begin{equation}
S_{liq}(T)=a \: T^{-2/5}+b.
\end{equation}
This corresponds to the assumption of a low temperature liquid energy behaviour $e \sim T^{3/5}$,
a theoretical prediction that seems to be well-followed in many simple liquids
\cite{RoTa}
and that we
find in very good agreement with our data, both in this case and in the
previously discussed data on the pressure.

In more detail, we performed a simulated annealing run of a system of $N=260$ particles,
in a cubic box with periodic boundary conditions, starting from
$\beta^{1/4}=0.02$ and performing
up to $2^{22}$ MC steps at each $\Delta \beta^{1/4}=0.02$, the maximum shift
$\delta_{max}$ allowed to each
particle in one step being chosen in order to get an acceptance $a \sim 0.5$.
The energy
and its fluctuation were measured in the last half of each run at a
given $\beta$ value while the
pressure was obtained from about 200 configurations in the same interval.
We take into account the small finite size effects by adding to the energy
and pressure the appropriate corrections evaluated analytically
by means of $\int_{\Lambda} v(r) d^3 r$, where $\Lambda$ is the volume
obtained as difference between the volume of the whole space and the volume of the box
containing the particles.
Just for decreasing the error on the evaluation of $S_{liq}$, we fit the very
high temperature
data on the energy, up to $\beta^{1/4}_0=0.1$, by using $\beta^{3/4}
e(\beta)= a \beta^{1/2}+b \beta^{1/4}+c$,
obtaining correspondingly $f(\beta_0)=4 a \beta_0^{3/4}/3+2b
\beta_0^{1/2}+4c\beta_0$, that turns out
to be perfectly compatible with the analytical value (i.e. we are still in the
region where no differences
are observable between numerical data and analytical approximations). The
integration is
subsequently performed by interpolating the simulation data with a standard numerical subroutine
in order to get a result which is independent of the integration interval.

In order to evaluate $S^{(a)}_{sol}$ and $S^{(b)}_{sol}$ we considered 16
different configurations
in the last half of the run at each $\beta$-value, while $S^{(c)}_{sol}$ was
measured from
the configurations obtained from these by 10000 MC steps at $T=0$
(starting from
 $\delta_{max}=0.05$ and decreasing it up to 0.005 during the run).

\begin{figure}[htpb]
\centerline{\epsfig{figure=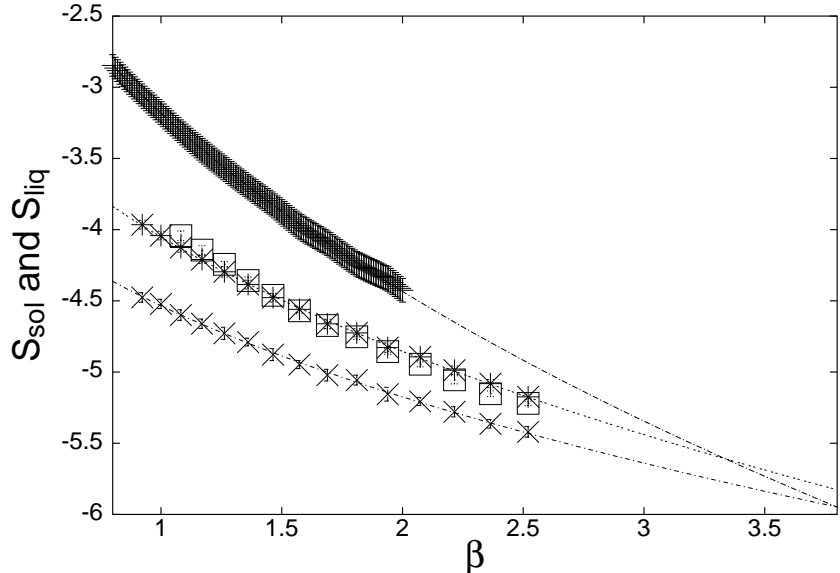,angle=270,width=11cm}}
\vspace{0cm}
\caption{The entropy of the liquid (+) and the different evaluations (see
text)
of the amorphous solid entropy, $S^{(a)}_{sol}$ ($\times$),
$S^{(b)}_{sol}$ ($\ast$) and
$S^{(c)}_{sol}$ ($\Box$), as functions of $\beta$. The lines are the
best fit to the power law, $S_{liq}=a \beta^{2/5}+b$.}
\label{entropie}
\end{figure}

In [fig. 8] we plot both $S_{liq}(\beta)$ and the different measurements of
$S_{sol}(\beta)$ considered. $S^{(b)}_{sol}$ and 
$S^{(c)}_{sol}$ are very close and give similar estimations of
$T_K \sim 0.3$, whereas $S^{(a)}_{sol}$ gives the slightly lower value
$T_K \sim 0.26$. It seems therefore reasonable to get the mean value 
$T_K \sim 0.28$ as our numerical evaluation of the Kauzmann temperature.

\begin{figure}[htbp]
\begin{center}
\leavevmode
\centerline{\epsfig{figure=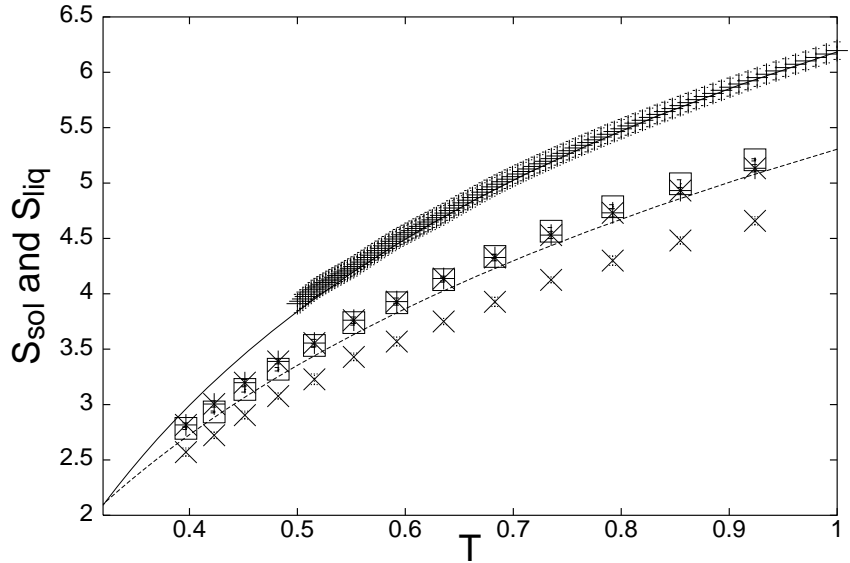,angle=270,width=11cm}}
\caption{Analytical $S_{liq}$ (upper line) compared with the
numerical values ($+$ points),
and
analytical $S_{sol}$ computed in the harmonic resummation scheme
(lower line) compared with the three
different numerical results, i.e. $S^{(a)}_{sol}$ ($\times$),
$S^{(b)}_{sol}$ ($\ast$), $S^{(c)}_{sol}$ ($\Box$).}
\end{center}
\label{configu}
\end{figure}

We note that this value of the Kauzmann temperature, to be compared with
 the analytical $T_K \simeq 0.32$, is close to the
previously discussed numerical estimation of $T_{GL}$. Our numerical data are
inadequate for clarifying this situation, which is intriguing and deserves a
 more careful study. Here we only suggest that
there could be a glass-gas phase coexistence in the
low temperature region.

Finally, [fig. 9] illustrates
the comparisons between
$S_{liq}$ and $S_{sol}$ as functions of the temperature $T$
computed analytically
(lines) and the same
quantities determined by our numerical simulations (points), as
discussed above. In this plot we added both the correct
normalization constant, giving $S=0$ when the volume of phase space is equal to
$h^3$, and the thermal contribution which had been neglected in the previous
figure. Obviously these terms do not affect the thermodynamical
glass transition. We used the physical parameters appropriate to argon
($\sigma_{++}=3.405 \: \AA$, $\epsilon_{++} / k_b =125.2 K$), plotting
the entropies per particle.

\noindent
The reliability of the approximations introduced is clarified by these
comparisons:

\begin{itemize}
\item
the numerical points of $S_{liq}$ are very close to the corresponding
analytical quantity computed by means of the interpolation
between HNC and MSA closures

\item
the three different ways to numerically evaluate $S_{sol}$
are in various agreement with the analytical curve, but it is
important to note that
in the low temperature region (below the dynamical temperature),
none of them is further than $\sim 10-20\%$ from the theoretical
prediction.

\end{itemize}

\end{subsection}

\section*{Acknowledgments}
We owe a lot to M.Mezard and we are very happy to thanks him.
We are also very grateful to W.Kob and F.Sciortino for their suggestions and
to A.Cavagna and I.Giardina for useful discussions.
B.C. would like to thank the Physics Department of Rome University
'La Sapienza', where this work was partially developed during her PhD.

\end{section}

\end{document}